\documentstyle{mn}
\input{epsf}
\newcommand\ie{{i.e. }}
\newcommand\eg{{e.g. }}
\newcommand\etal{{et al. }}

\newcommand\apj{{ApJ}}
\newcommand\mnras{{MNRAS}}
\newcommand\apjs{{ApJS}}
\newcommand\apjl{{ApJL}}
\newcommand\aj{{AJ}}

\begin{document}
\title{Cosmological Evolution of Supergiant Star-Forming Clouds}
\author{Melinda L. Weil $^{1,2}$ \& Ralph E. Pudritz $^{1}$\\
$^1$Department of Physics and Astronomy, McMaster University,
1280 Main Street W., Hamilton, Ontario, L8S 4M1 Canada\\
$^2$City University of New York, New York}
\maketitle

\begin{abstract}

In an exploration of the birthplaces of globular clusters, 
we present a careful examination of the formation of
self-gravitating gas clouds within assembling dark matter
haloes in a hierarchical cosmological model.  Our high-resolution
smoothed particle hydrodynamical simulations 
are designed to determine whether or not hypothesized
supergiant molecular clouds (SGMCs) form and, if they do, 
to determine their physical
properties and mass spectra.  It was suggested in earlier work
that clouds with a median mass of several
$10^8 M_{\sun}$ are expected to assemble during the formation of a galaxy, 
and that 
globular clusters form within these SGMCs. 
Our simulations show that clouds with the predicted properties are indeed
produced as smaller clouds collide and agglomerate
within the merging dark matter haloes of
our cosmological model.  We find that the mass spectrum of
these clouds obeys the same power-law form, $dN/dM \propto M^{-1.7 \pm 0.1}$,
observed for globular clusters, molecular clouds, 
and their internal clumps in galaxies,
and predicted for the supergiant clouds in which
globular clusters may form.   We follow the evolution and 
physical properties of gas clouds
within small dark matter haloes 
up to $z = 1$, after which prolific star formation is expected
to occur.  Finally, we discuss how our results  
may lead to more physically motivated "rules" for star formation
in cosmological simulations of galaxy formation.

\end{abstract}
\section{Introduction}

The picture of galaxy formation as a hierarchical clustering process
(\eg White \& Rees 1978; Fall \& Efstathiou 1980) has gained
considerable observational support over the last decade (\eg
Giavalisco, Steidel, \& Macchetto 1996; Steidel \etal 1998; Abraham
\etal 1999).  It is thought that evolving dark matter haloes set the
stage for the subsequent gas cooling and star formation that
ultimately result in the plethora of galaxies we see in the present
epoch.  However, the lack of a detailed physical theory of how stars
form in early, evolving environments remains a
pressing problem. For example, when gas dynamics and star formation are 
ignored, distributions of hundreds of dark matter 
satellites around larger galaxies have been produced in numerical simulations (Klypin \etal 1999; Moore \etal 1999) whereas only tens of satellites are 
observed.  In addition, simulations of galaxy
formation in the context of Cold Dark Matter (CDM) dominated universes
assume that thermal cooling-function prescriptions can be used
to predict how and when
gas converts to stars. 
If it is supposed that
the bulk of the cooled gas goes directly into star formation,  
then the inescapable high cooling rate that occurs in gas
within the growing dark matter haloes when the universe was much
smaller and denser leads directly to the so-called
`star formation catastrophe'. 
Far too high a proportion of baryons end up in stars in such simulations
compared to the observations.
Simulations also repeatedly fail to produce Milky Way
types of galaxies because their gas condenses too quickly to be torqued
up to the observed angular momentum of gas in spirals (Navarro \& 
Steinmetz 1997; Weil, Eke, \& Efstathiou 1998; Eke, Efstathiou, \& 
Wright 2000, hereafter EEW).  It is hoped that the remedy for these
ills lies in the addition of the feedback of star formation upon the
gas. Vigorous supernova activity may heat up the gas enough for it to
remain fairly diffuse and extended (\eg McKee \& Ostriker 1977, McKee 1995).

The purpose of the present paper is to apply the insight of
star formation processes in the current epoch to 
help solve this problem.
The observations of star formation in nearby molecular clouds 
and nearby extragalactic systems clearly show that
the prescriptions for star formation imposed in the 
cosmological models are too simple.
Firstly, self-gravitating (molecular) clouds are not supported
by thermal pressure, but rather by a combination of turbulent
and magnetic pressure (McKee \etal 1993; Pudritz 2000).
Thus, simple thermal cooling time arguments are
irrelevant for predicting how molecular clouds convert some of
their mass into stars.
Secondly, stars form in clusters within clumpy subregions of molecular
clouds, and the total mass of these embedded clumps is only a 
few percent of the mass of the
entire cloud. Hence, star formation proceeds 
only in sub regions within molecular clouds 
that are sufficiently self-gravitating, or 
equivalently, are at sufficiently high pressure.  Thirdly, the  
mass spectra of molecular clouds and that  
of their clumps have been 
found to obey an approximate power law scaling
\begin{equation}
dN / dM \propto M^{-1.7 \pm 0.1}
\end{equation}
(eg. Blitz 1993, Williams et. al. 2000).
The fact that stars form within such special sub-structures
in the self-gravitating clouds almost
certainly has important implications for the formulation of more
physically motivated ``rules'' for star formation.

Do characteristics of galactic star formation in the present epoch
carry over to the epoch of galaxy assembly?
Strong indications that the answer is yes come from the old halo globular
cluster systems that have been studied around more than
100 galaxies (see review; Harris 2000). Harris and Pudritz (1994, hereafter HP)
showed that the mass spectrum characterizing globular cluster
systems around a variety of galaxies
has nearly the identical form as the mass spectrum
for local giant molecular clouds (GMCs) and their
clumps. 
Because the efficiency
of conversion of clump gas to stars in bound clusters must be high,
the mass spectrum of bound clusters should reflect
that of their progenitor clump spectrum. 
Observations show that globular clusters more massive than $10^5 M_{\sun}$
and more than a few kpc from the centers of galaxies survive evolution
in their host galaxies relatively intact (\eg Harris, Harris, \& 
McLaughlin 1999).  This is born out by semi-analytical studies of the tidal
evaporation and disruption of globular clusters in isothermal dark matter
haloes for a very wide range of cluster masses, concentrations, and orbital
parameters (\eg Capriotti \& Hawley 1996).  Such calculations show that 
clusters in the mass range $10^5$ to $10^7 M_{\sun}$ can survive for a
Hubble time because they are massive enough to avoid evaporation but
not too massive to undergo rapid orbital decay though dynamical friction.
Thus,
the mass spectrum of the more massive globular clusters around
any galaxy essentially preserves the mass spectrum
of its  progenitor clumps. The numerical model of 
McLaughlin and Pudritz (1996, hereafter MP)
showed that such clump mass spectra arise naturally through
an agglomeration/fragmentation process, wherein clumps and clouds build up
by the collision and merger of smaller ones within
any given dark matter halo, and are destroyed as a consequence of star
formation within them.  

Globular clusters also provide empirical  
guidance as to how stars formed during galaxy assembly in a wide variety of
systems and environments.  The globular cluster 
luminosity function or, equivalently,
mass spectrum, around galaxies has been measured
in a variety of galaxy types such as dwarf ellipticals (\eg Durrell
\etal 1996), as well as spirals and ellipticals. The mass spectra 
are found to be identical
to equation (1) within the errors.  The mass spectra of  
globular cluster
systems is also independent of the metallicity of the host galaxy (\eg
MP).  Moreover, globular clusters are observed to form today in galaxies that
have abundant supplies of gas (\eg Holtzman \etal 1992; Whitmore \& 
Schweizer 1995; Zepf \etal 1995; Schweizer \etal 1996; Whitmore \etal
1999).  Thus, globular cluster formation has a universal
character which requires only a sufficient supply of gas,
and is independent of host galaxy conditions and details about
gas metallicity.

Previous work on this model for cluster formation focussed on
processes occurring within some given dark matter halo, and did not
fully embed these ideas within a full cosmological framework.  Thus,
HP and MP constructed a model in which globular clusters form within
massive clumps of $10^5 - 10^6 M_{\sun}$ that are embedded within
their yet larger host, supergiant self-gravitating (partially
molecular) clouds, denoted SGMCs.  The hypothetical SGMC mass range --
$10^7 - 10^9 M_{\sun}$ -- is in fact observed for molecular gas within
merging or starburst galaxies (\eg Wilson \etal 2000).  In the model,
the mass spectrum of the SGMCs (and their substructure - the clumps )
arises from the agglomeration of gas clouds by cloud-cloud collisions
within the evolving dark matter haloes. The model envisages gas
processes within any given, static dark matter halo but does not
include the mergers of such haloes that arise in any hierarchical
model for galaxy formation. Thus, globular cluster formation is
predicted to occur much as star cluster formation does in molecular
clouds at the present day, with the only difference being that huge
gas reservoirs were available (through gas infall or ongoing mergers)
at these higher redshifts in comparison with current interstellar
media.

In the present paper, we undertake, for the first time, a careful
examination of the formation of self-gravitating clouds that build up
during the assembly dark matter haloes in hierarchical cosmological
models.  Our purpose is to determine whether or not supergiant clouds
form, to measure their mass spectra, and to determine how they evolve
from early redshifts to $z = 1$.  Our high resolution simulations
focus, in particular, on the study of these progenitor supergiant
clouds before the "star formation catastrophe" occurs (typically at
redshifts $z \simeq 1$).  We specifically designed our simulations to
have the resolution needed to study the formation of ensembles of
individual, self-gravitating (and partially molecular) clouds that are
the actual seats of star formation.  The idea is to test and survey
the initial conditions for star formation, {\it before} problems such
as the star formation catastrophe become endemic.  This procedure will
allow us to test current models of star and globular cluster formation
on cosmological scales. It will also ultimately lead to the
development of more physically motivated rules for stellar formation
in cosmology.

Our results clearly show that SGMCs are commonly formed
within evolving dark matter haloes in our first CDM cosmological
model, and that they have the
the mass spectra and cloud properties predicted
for clouds capable of forming globular clusters.
We discuss our numerical methods in \S 2, and present
our results in \S 3. Implications of our work for star formation
during galaxy assembly and, hence, for globular cluster formation 
follow in \S 4.

\section{Numerical Methods}

We investigate the formation and evolution of
star forming clouds in one of the several viable,
spatially flat, CDM cosmological
models (Thomas \etal 1998), known as $\tau$CDM.
Structure formation in dark matter and
gas is followed using smoothed particle hydrodynamical simulations (SPH). 
We evaluate the formation of supergiant star-forming clouds during
the evolution of small density perturbations initially seeded in an
expanding universe.

In order to attain the mass resolution required to
identify massive, self-gravitating clouds and their substructure,
we choose small dark matter haloes
from a large cosmological simulation which has been evolved to
the present day ($z = 0$).  After tracing the dark matter haloes
back to redshift $z=24$,
their resolution is improved by increasing the particle numbers by
factors of hundreds and adding an equal
number of gas particles.  The volume outside the high resolution
region is represented by high-mass, low resolution particles to retain the
full gravitational characteristics of the original model.
The models are then evolved from $z=24$ to $z=0$.
Section 2.1 presents properties of the large
cosmological simulation and the method of producing high
resolution models. The properties of the dark matter plus gas simulations
the code used to evolve the high resolution models
are presented in \S 2.2.

\subsection{Cosmological Model}

A dark matter only cosmological simulation is used to generate
initial conditions for the high-resolution dark matter plus gas
simulations described below.  The cosmological model is one of two spatially
flat CDM simulations by EEW.

The $\tau$CDM cosmology used in this paper has a matter density
parameter $\Omega_m=1$, zero cosmological constant, and a Hubble
constant $h=0.65$ in units of $100{\rm km}{\rm s}^{-1}{\rm Mpc}^{-1}$.
The power spectrum that generates the scale-free initial density
fluctuations for these simulations has a shape parameter of
$\Gamma=0.2$, which fits observed large-scale correlations in galaxy
and cluster surveys (Efstathiou, Bond, \& White 1992). This specific
cosmological model could be relevant to the situation in which the
late decay of $\tau$ neutrinos produces an additional background of
relativistic $e^-$ and $\mu$ neutrinos.  This has the effect of
delaying the onset of matter domination.  The other large cosmological
simulation produced by EEW is a $\Lambda$CDM model with $\Omega_m=0.3$
and $\Omega_{\Lambda}=0.7$.  The ages for the $\tau$CDM and
$\Lambda$CDM cosmological models are $10.0$ Gyr and $14.5$ Gyr,
respectively. Due to its shorter evolution time, we chose to explore
the formation of star-forming clouds in the $\tau$CDM first.  
No significant differences in the physical processes related 
to the evolution of SGMCs are expected for 
other cosmologies, although the precise time when SGMCs first 
turn up will of course be different (see \S4 for further discussion).  
The results on the formation of star-forming clouds in $\Lambda$CDM models
will be reported in a subsequent paper.

The large dark matter simulation represents a cubical volume of the
universe of length $L=32.5 h^{-1}$ Mpc with $N=128^3$ dark matter
particles.  Figure 1 shows a 2-Mpc deep slice of the simulation at $z=0$.
EEW identified virialized haloes using a group-finding method, locating
over 2000 haloes with
particle numbers ranging between about 40 and 30,000.
The largest dark matter potential wells are easily seen at the nodal
intersections of the filaments.  One of the small haloes we chose to
resimulate at high resolution is identified by the arrow near the top
middle of the figure.  This halo contains 65 dark matter particles and
has a mass $M=4.5\times 10^{11} M_{\sun}$.

\begin{figure*}
\centering
\centerline {\epsfxsize=15.0cm \epsfbox[18 144 592 718]
{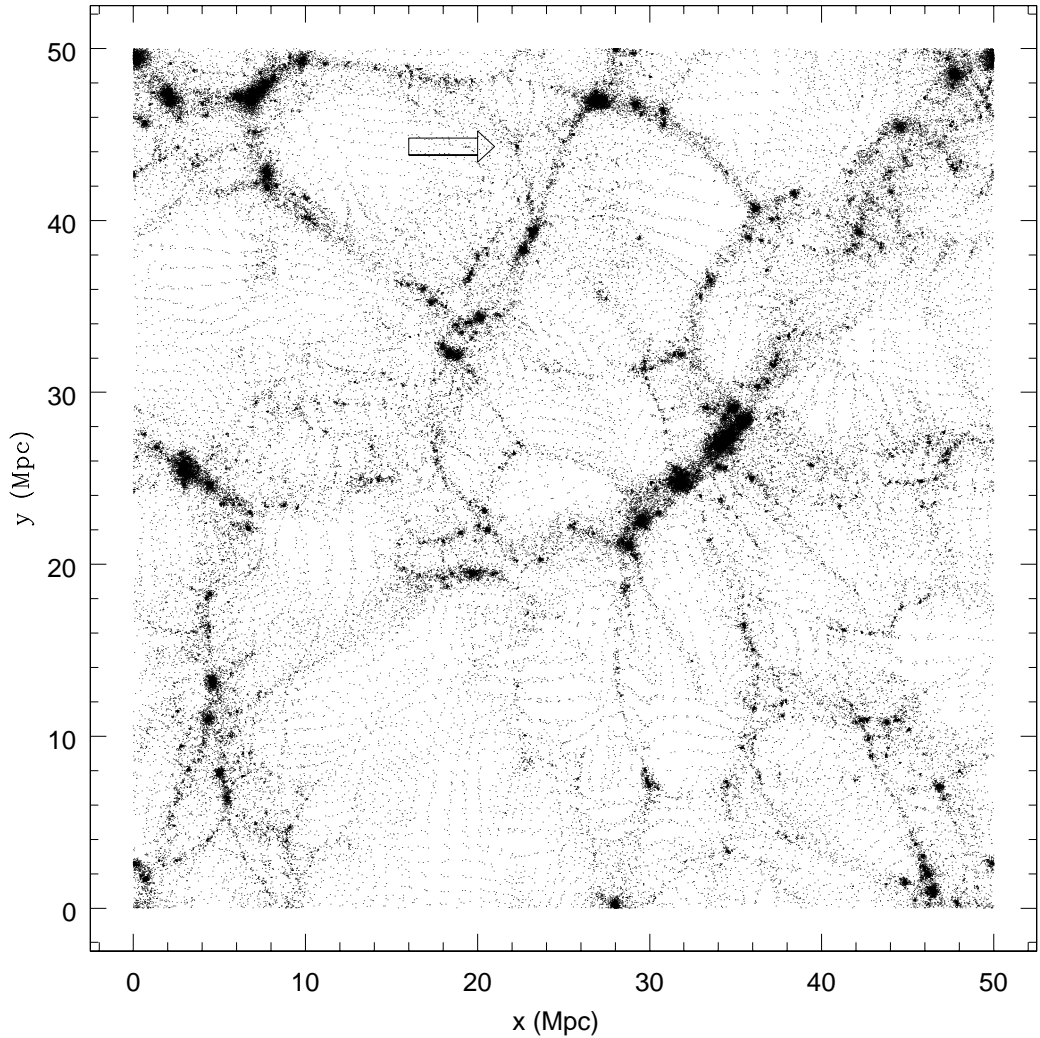}}
\caption{$2$-Mpc deep slice of the large, dark matter only
cosmological simulation of EEW.  The arrow near the top middle
points to the small halo which is resimulated at high resolution and shown
in Figures 2 and 3.}
\label{fig:figure1}
\end{figure*}

Haloes to be resimulated at higher resolution are chosen using a set
of balanced criteria.  Firstly, the mass of each halo is required to be
relatively small because computing time increases as the number of particles
increases.  We represent the high resolution region of the simulation
by $N_{dm}=34^3$ dark matter particles and $N_g=34^3$ gas particles.  The
surrounding low resolution region contains approximately $5000$ dark
matter particles representing the gravitational forces due the the
mass outside the high resolution region.  In order to achieve the mass
resolution necessary to examine the structure supergiant star-forming
clouds, the masses of the small haloes are limited to approximately
$5.0 \times 10^{11} M_{\sun}$.  Secondly, we choose haloes comprised of
enough particles to be physically believable structural entities. The
large cosmological simulation is searched for haloes containing
$\approx 60$ particles. When expanded to high resolution, the haloes
contain gas particles with masses between $2$ and $5 \times 10^6
M_{\sun}$.  Thus, while we have sufficient numerical resolution
to study the formation of SGMCs, our simulations do not yet allow
us to resolve the individual clumps within these clouds that
will be the actual sites for the formation of star clusters.
In these simulations, the baryon density parameter
$\Omega_b=0.06$; thus, 94\% of the mass is contained in the dark
matter particles. Finally, haloes are required to be well-separated
from any other large mass condensations so that computational power
will not be exerted upon nearby regions of little interest.

The high resolution initial conditions are created using the method
described by Weil \etal (1998).  After a halo is chosen at $z=0$, all particles
within a spherical volume surrounding it are located at $z=24$, the
redshift at which the cosmological simulations commence.  A cube just large
to contain all the particles is constructed. The length of the cube is
typically a few Mpc.  Populating the high resolution cube requires
displacing particles from a regular lattice in order to represent the spatial
fluctuations in density of the cosmology.
In order to correctly populate their finer volumes, these high resolution 
particle displacements require
additional calculations that add power above the Nyquist
frequency of the original simulation.  The higher mass, low resolution region
particles are then added in a spherical distribution.

\subsection{Numerical Simulations}

In order to allow statistical analysis of the properties and evolution
of star-forming clouds, seven different dark matter
haloes were chosen for high
resolution resimulation.   We studied the evolution of
each of these systems, which we designate by distinct
Run numbers: 1 -- 7. Each Run was evolved using the
TREESPH code (Hernquist \& Katz 1989).  This code treats both dark
matter, which is purely gravitational, and gas, which includes
hydrodynamical and radiative physics (for details, see Weil \etal 1998).

The gravitational forces between particles are computed using a
hierarchical tree algorithm (Barnes \& Hut 1986).  Distant particles
are grouped into hierarchical cells in which the potential is
approximated using a truncated multipole series.   The forces on
each particle are updated by examining each cell in the tree.
The ratio of the size of the cell to its distance from the particle
is compared to a defined tolerance parameter to determine whether
the cell should be subdivided or the forces due to it calculated as
a whole.
In this way,
computational time is reduced by avoiding explicit two-body force
calculations for remote particles, while an accuracy of $\sim 0.1\%$
is maintained by a tolerance parameter of $0.6$.  

The gravitational
potential is softened using a cubic spline in order to reduce two-body
relaxation. The softening lengths assigned to the particles define the
length resolution of the code.  For dark matter particles, the
softening $\epsilon _{dm}$ is always between $2$ and $3$ kpc.  For the
gas particles, with masses more than a decade smaller than the dark
matter, $\epsilon _{g}$ is between $0.6$ and $1.0$ kpc.
This gaseous softening length imposes another restriction
on the size of clouds that we can physically resolve, namely,
nothing less than several kpc.  Given that the HP model suggests
a fiducial value of one kpc for the median SGMC in the distribution,
we caution that our simulations will inevitably produce fluffier, lower
density clouds by virtue of the limitations imposed
by the softening length.  Detailed analysis of cloud structure will require 
future, higher resolution simulations.

Hydrodynamical interactions between gas particles are treated
using smoothed particle hydrodynamics (Lucy 1977, Gingold \& 
Monaghan 1977).  The gas is partitioned into self-gravitating fluid
elements that evolve according to Lagrangian hydrodynamic
conservation laws.  Explicitly, pressure gradients and the
viscosity of the medium are calculated using conservation of momentum:
\begin{equation}
    {{d{\bf r}}\over {dt}}={\bf v} , \label{eq:r5}
\end{equation}
and
\begin{equation}
    {{d{\bf v}}\over {dt}} = -{1 \over {\rho_{gas} }} \nabla P +
     {\bf a}^{visc} - \nabla \phi({\bf r}) , \label{eq:r6}
\end{equation}
where ${\bf r}$ and ${\bf v}$ are vector positions and velocities,
$\rho_{gas}$ is the gas density, $P$ is the pressure, ${\bf a}^{visc}$
comprises the acceleration due to viscosity, and $\phi$ is the
gravitational potential.  The equation of state if that of an ideal
gas; stability during collisional shocks is maintained by an
artificial viscosity (Navarro \& Steinmetz 1997).

In order to treat the evolution of the thermal energy, radiative
cooling is computed from the ionization, recombination, and cooling
rates of Black (1981, Table 3) with modifications by Cen (1992).
Analytic approximations to observed rates of collisional ionization,
recombination, collisional excitation, bremmstrahlung, and Compton
heating and cooling are constructed for H I, H II, He I, He II, and
He III.  Because we ignore star formation, rather than initializing and
evolving metallicities for the gas particles, primordial gas is
assumed to have a hydrogen mass fraction of $X=0.76$ and a mass helium
fraction of $Y=0.24$. As already noted, observations of the IMF and
the overall characteristics of the mass functions of globular clusters
indicate they are independent of the metallicity of the gas out of
which they formed.

The local quantities appearing in the hydrodynamical equations are
computed by interpolating between gas particles (Monaghan 1985,
Steinmetz \& M\"uller 1993).  Individual smoothing lengths are
calculated for each particle in order to ensure that the values for
each particle are calculated with reference to $25$ to $45$ neighbors.
In addition to individual smoothing lengths, individual timesteps
are permitted for each particle.
The largest time step for these TREESPH simulations
is $\Delta t = 2.5 \times 10^6$ years; the smallest number of timesteps
in which a particle can reach $z=0$ is $N_{step}=4000$.
Owing to the need to satisfy local stability as required
by the Courant condition, particles in the highest density areas
are allowed to reduce the size of their timesteps by up to a factor of 32.  In
simulations with a large range of mass inhomogeneities, this
adaptivity in space and time permits areas of high density to be
resolved to an arbitrary resolution.

Our simulations deliberately avoid the implementation
of any star formation rule.  This enables us to
study the properties of ensembles of self-gravitating
clouds that build up before any star formation catastrophe --
predicted by a naive interpretation of gaseous thermal cooling
processes -- could be important.
This assumption implies that we cannot chart the evolution of the
metallicity of these early systems before $z=1$.  We therefore refrain
from speculating about the observed metallicities of globular
cluster systems, or any observed trends of this kind (\eg age-metallicity
relations) in any given galaxy, and how these may come to be.  Without
the high spatial and temporal resolution necessary to resolve and follow
the star formation in clumps within SGMCs, such discussion would be premature.
The second, and related, limitation
is that the destruction of clouds due to star formation is
not taken into account. The proper treatment of cloud
destruction critically depends however, upon the incorporation
of correct star formation physics. Both of these limitations
will be relaxed in subsequent work. 

The quantities discussed thus far refer to the fiducial initial
conditions for our simulations.  However, each of the seven haloes
was first resimulated at ``low resolution.'' The low resolution
simulations were assigned particle numbers for dark matter and for
gas
of $N_{low}=22^3$ each in comparison to the $N_{high}=34^3$ each
of the high resolution simulations.  Simulations at this resolution
require less than one day to compute on a Sun Ultra 10 workstation.
Each halo was evolved from $z=24$ to $z=0$ in low resolution; and
was examined for any aberrations that would bias the results. Systems
that were in the process of merging or consisted of several 
subsystems at $z=0$ were excluded.
Subsequently, each halo was evolved in high resolution from $z=24$ to $z=1$.

Figures 2 and 3 show the evolution of  dark matter and gas,
respectively, during Run 2.  The length of
each panel is 2 Mpc in comoving coordinates and the redshifts are
marked at the upper right.  In this hierarchical cosmology,
several dark matter potential wells, into which star-forming gas falls,
have formed by $z=5$.  The gas is more centrally concentrated than the
dark matter, forming small, denser clumps, surrounded by more diffuse
gas. If the first epoch of efficient star formation occurs at $z=5$ in
our adopted $\tau$CDM  cosmology, then the present age of the
first objects is 9.7 Gyrs.  In comparison,
if stars first form efficiently at this time in a standard CDM
cosmology, their age is $\approx 11$ Gyrs.
Even at $z=1$, several gas condensations, which will eventually
merge to form a small galaxy, are visible.  The present age of stars formed at this time
is 6.5 Gyrs in $\tau$CDM and $\approx 8$ Gyrs in standard CDM.
We discuss the significance of these ages in \S 4.

\section{Results}

In order to construct a detailed theory for the formation of globular cluster systems
(GCS) and, more generally, for star formation within hierarchical cosmologies,
the results of simulations need to be compared to observational
quantities and models. If stars do indeed form at early epochs in the high pressure
clumps within massive gas clouds (as they do in the present
epoch), then high resolution cosmological simulations of
galaxies should enable the identification of these supergiant clouds.
Herein, bound gas masses that represent the star-forming clouds within
Runs 1 - 7 are found and examined.
The time evolution and statistical and individual properties of these supergiant,
star-forming clouds are also investigated. 

\begin{figure*}
\centering
\centerline{\epsfxsize=15.0cm \epsfbox[18 144 592 718]
{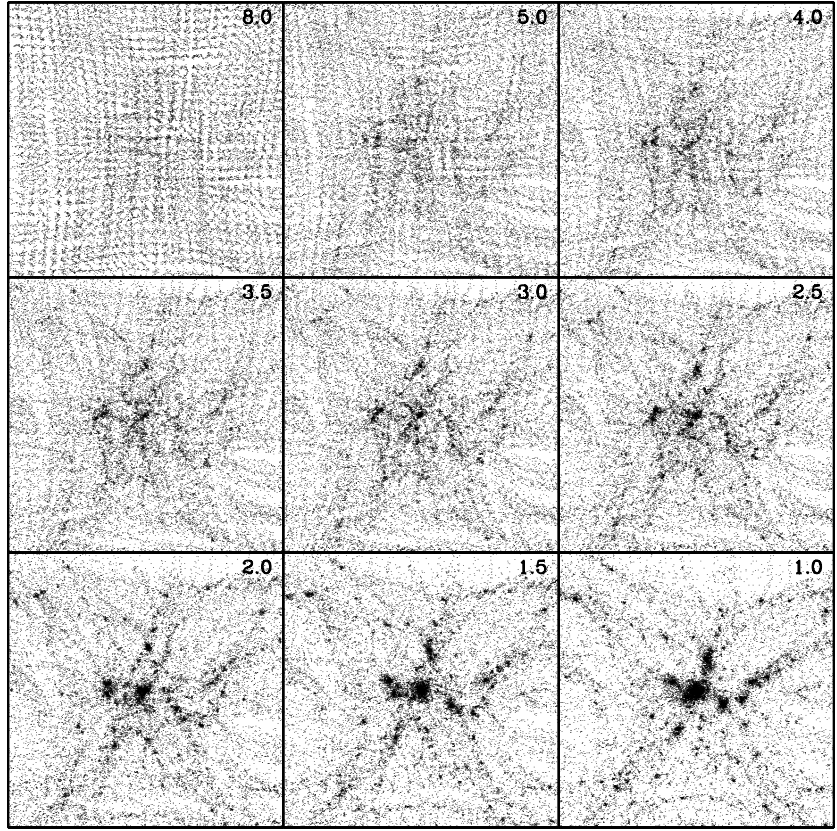}}
\caption{Time evolution of Run 2 dark matter.
Each panel measures 2 Mpc in length in comoving coordinates.  Redshift
is shown in the upper right hand corner of each panel.}
\label{fig:figure2a}
\end{figure*}
\begin{figure*}
\centering
\centerline{\epsfxsize=15.0cm \epsfbox[18 144 592 718]
{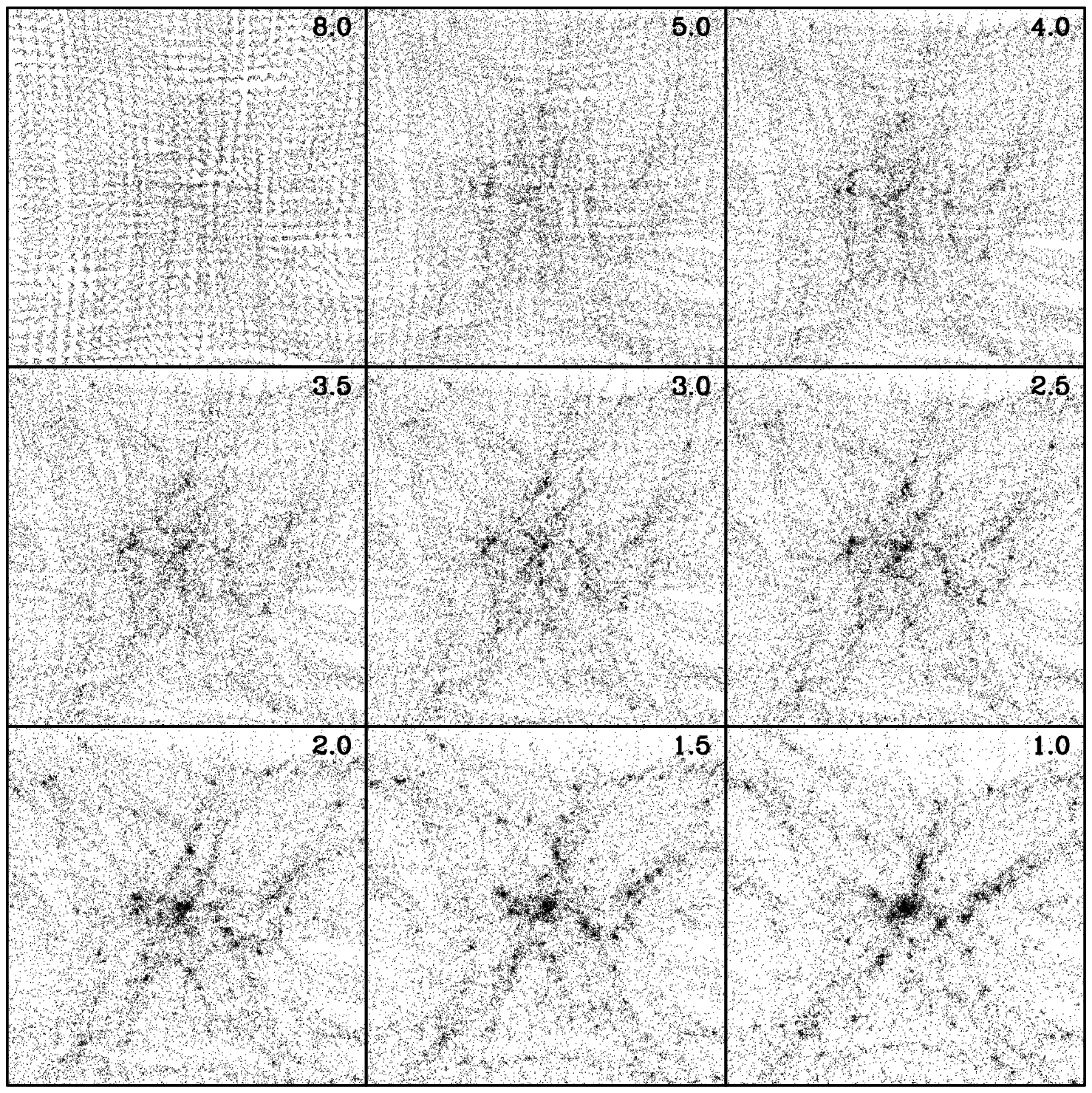}}
\caption{Same as Figure 2 except gas is shown.}
\label{fig:figure2b}
\end{figure*}

Our main goal was to identify sufficiently massive bound gas clouds formed in high resolution simulations such as that shown in Figures 2 and 3.
Two methods of identifying clumps of particles were applied to
outputs at several redshifts of Runs 1 -- 7.  Because the clumps interact
with one another and can be irregular, overlapping, and not in
dynamical equilibrium, we require an algorithm which can uniquely locate
separate clumps and ensure that the systems are bound.

The spherical overdensity algorithm (SO; Lacey \& Cole 1994) grows spheres
around density maxima until they reached a defined density contrast with
the background.  The overdensity $\Delta=178$
corresponds to virialized regions within a uniform collapsing
sphere. The friends-of-friends algorithm (FOF; Davis \etal 1985) uses a
linking method by which particles are identified as members of a
clump if they are closer to the nearest member of the clump than a
defined `link length.'  In essence, the link length determines a local
density criterium.  The results of the two clump-identifying algorithms
supports the Lacey and Cole (1994) observation that small clumps
are sometimes merged with larger ones for the fiducial
parameterizations. The fiducial FOF link length is 0.2 times the mean interparticle
separation. Herein, however, the typical link length is defined to
be about a factor of two less than that, in order to separately
identify small clumps and to avoid including unbound particles.  It is
likely that, for many clumps, several outlying particles are not included
as members; however, the mass contribution of these few particles is
negligible in most cases.

For the several output redshifts of Runs 1- 7, all clumps with
particle numbers $N_{c,g}>15$ and $N_{c,dm}>30$ for gas and dark
matter, respectively, are located.  Subsequently, only those
clumps whose particles are enclosed within the virial radius of the small
galaxy at z=0 are retained.  Those clumps with gas masses greater than a few
times $10^7 M_{\sun}$ are considered to be supergiant star-forming
clouds.  We apply the clump-finding algorithm to three different
types of bound objects, namely, to gas plus dark matter agglomerates, to
pure gas clouds and, finally, to pure dark matter haloes.

\subsection{Cloud Populations}

The epochs at which star-forming clouds
appear are investigated for Runs 1 -- 7.  At our mass resolution 
limit of a few times $10^6 M_{\sun}$, by
$z=5$ each of Runs 1 -- 7 produces 2 to 10 gas clouds with masses
of several times $10^7$ to several times $10^8 M_{\sun}$. These will
eventually become part of the small galaxy.  Figure 4 shows the
evolution of supergiant star-forming clouds located by FOF at
eight redshifts for Run 2.  Each of these gas clouds consists of
particles identified to be contained within the virial radius of the small
galaxy at $z=0$.  Each panel measures 1 Mpc in length.
In the bottom right panel, the gas particles at
$z=0$ are shown.  The circle outlines the area within the virial
radius of the galaxy, which encloses $4.4 \times 10^{11} M_{\sun}$
(including dark matter and gas).  At $z=1$, the clouds contain
$1.2 \times 10^{10} M_{\sun}$ gas.

\begin{figure*}
\centering
\centerline {\epsfxsize=15.0cm \epsfbox[18 144 592 718]
{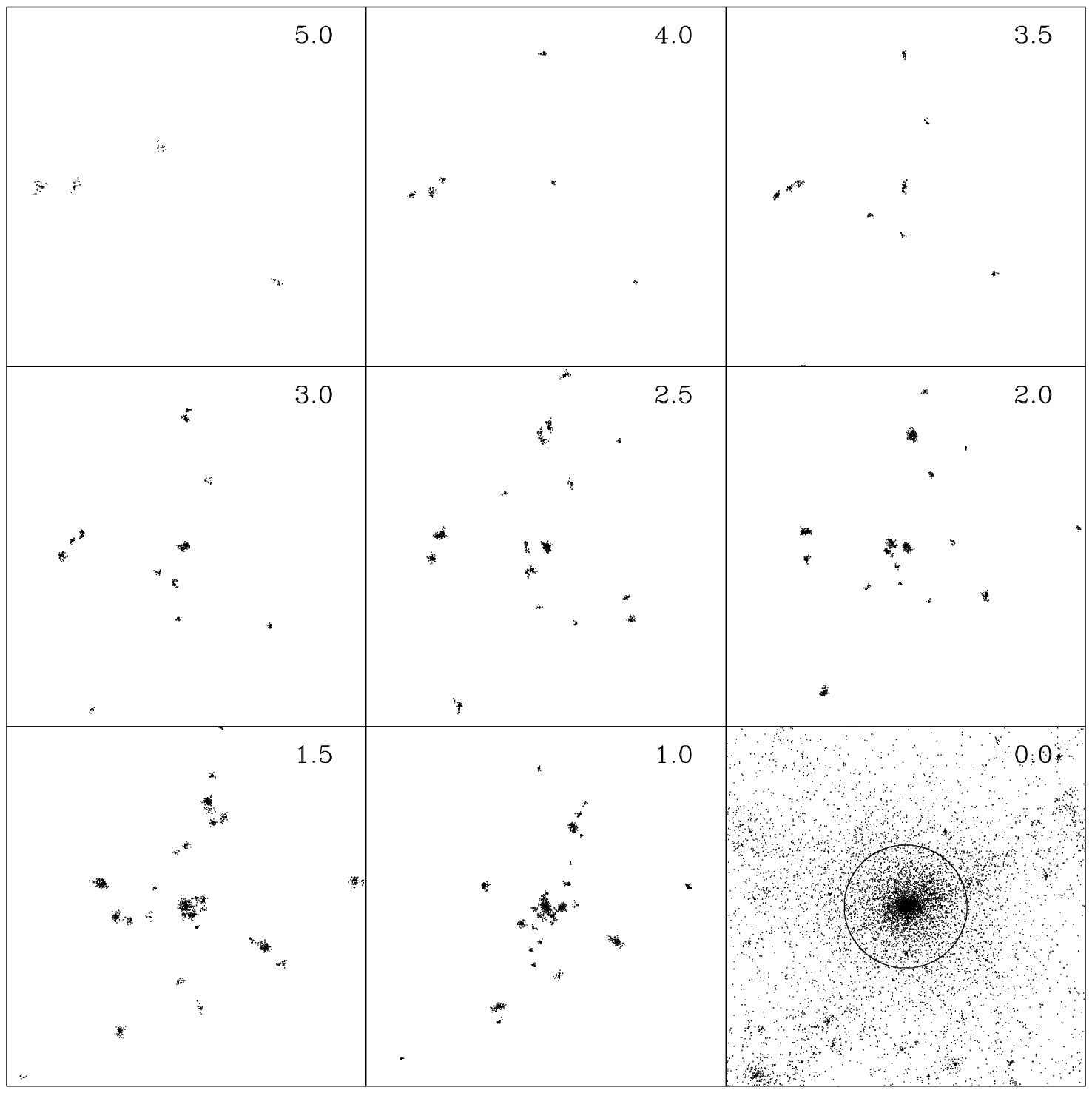}}
\caption{Bound gas clouds at eight redshifts between $z=5$ and $z=1$ for
Run 2. The clouds merge eventually to become part of the small galaxy
shown at $z=0$. The circle outlines the area within the virial radius.
Each panel measures 1 Mpc in length in comoving coordinates.
}
\label{fig:figure3}
\end{figure*}

Figure 4 shows the bound gas clouds at eight different redshifts,
and their final product at $z = 0$.
The number of star-forming clouds increases with
redshift as the density perturbations of the hierarchical
cosmology grow large enough to contain bound material.  Although initially
well-separated in space, clouds flow along the filaments visible
in Figures 2 and 3 towards the dominant density concentration, piling up in
the center as they merge to form a galaxy.  The clouds appear
irregular in shape throughout all redshifts.  It is clear from this
figure that clouds are growing by cloud-cloud collisions at the
same time as the galaxy is building up through a hierarchical
merger process.

One can clearly see the collision and agglomeration of clouds
in Figure 4.  At  $z = 3.0$, we see 3 distinct
clouds in middle of the left side that collide,
leaving two clouds at that position in the panel at $z=2.5$.
As another example, the two amorphous clouds in the
upper middle part of the $z= 2.5$ panel collide
to produce the larger cloud at the same position 
in the $z=2.0$ panel. Finally, the group of clouds in the middle of the
panel at $z=2.5$ collides and agglomerates by $z = 1.5$.
These, and other details clearly show that
gas agglomerates into more massive structures by cloud-cloud collisions.
Gas is brought together in two ways: 1)
it falls into an individual dark halo and 
collides with other clouds in that halo, 
and 2) gas in a given halo collides with the fresh gas contained in
other small dark matter haloes that are merging.

Figure 5 shows the central region of Run 2 at $z=1$.  The left
panel shows the gas and the right panel shows the bound gas clouds.  
The colors of the clouds are non-physical;
all the gas particles that belong to a distinct cloud are represented by a single color.
The dark matter particles are not included in this view. As a rule, clouds consist of a
dense central core, which contains most of the mass, surrounded by
more diffuse gas. We caution that we cannot
further resolve the expected substructure within
our clouds, because of the limitations imposed by our resolution.  In
the same way that the halo in the original cosmological simulation
of Figure 1 is essentially featureless, these clouds, in which the
mass of each gas particle is on the order of the mass of a globular
cluster, are not complex.  However, if we examine the ensemble of
clouds in the center of Figure 5 as a whole, it has the appearance
of an amorphous agglomerate containing several high
density cores.  If the mass resolution for Runs 1 -- 7 is further
increased, the individual clouds would likely show further
geometrical complexity.

\begin{figure*}
\centering
\centerline {\epsfxsize=15.0cm \epsfbox[18 144 592 718]
{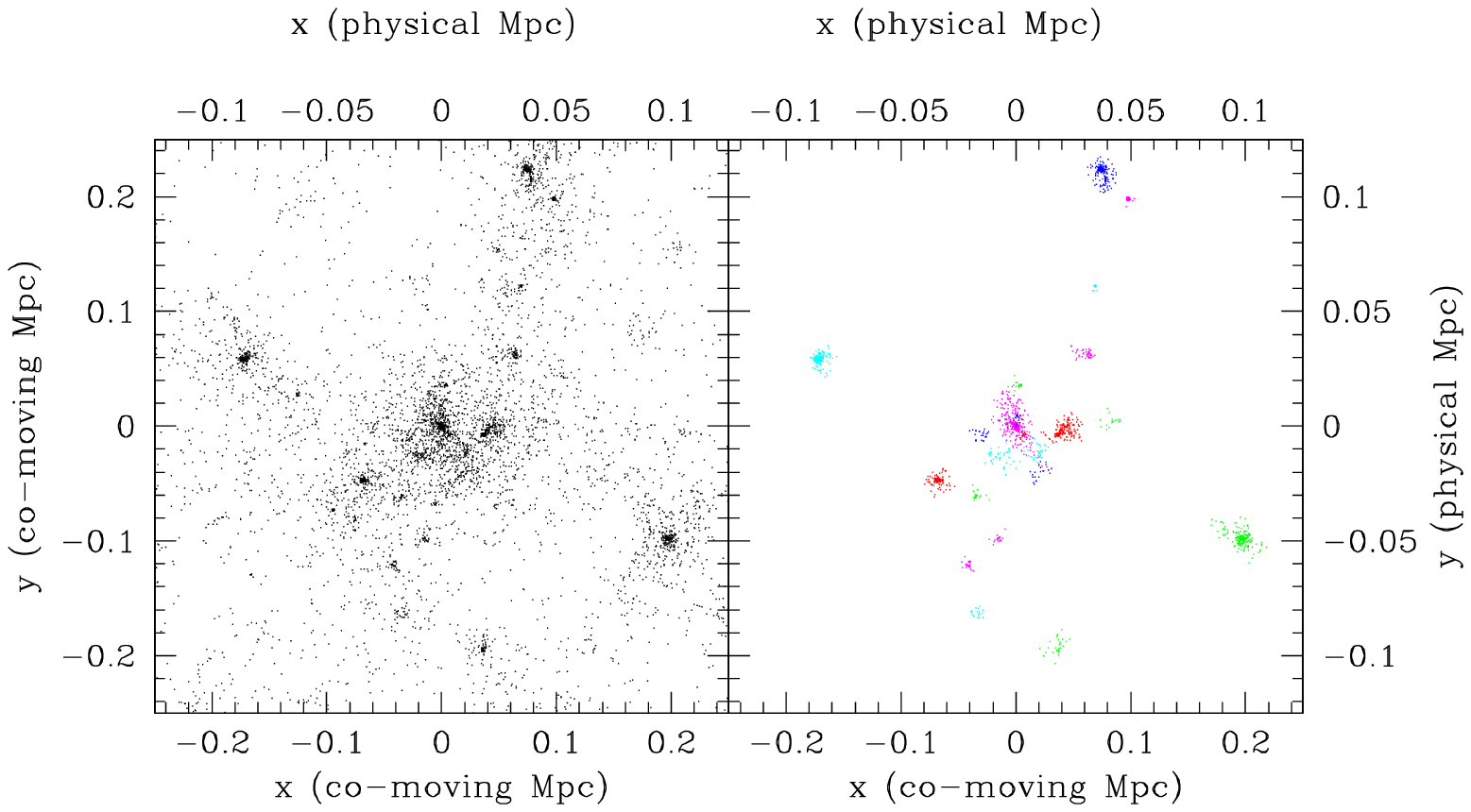}}
\caption{Run 2 at $z=1$.  The left panel shows the
gas in the central region and the right panel shows the bound gas
clouds. Each panel measures 0.5 Mpc in length in comoving coordinates, as
labeled on the left and bottom axes.  Also shown are physical
Mpc on the right and top axes.
}
\label{fig:figure4}
\end{figure*}

Each of the seven separate evolving haloes is accompanied
by a spectrum of masses for both the gas clouds and the dark matter.
Because the number of clouds in each separate Run can be small,
the collections of bound objects in each of Runs 1 -- 7 are added
together at each redshift in order to give a larger statistical sampling.
At a redshift of $z=5$, the haloes consist of between 2 and 11 clouds each.
At intermediate redshifts, most haloes consist of about 15 clouds but one
contains only 3 and another contains about 30.  At $z=1$, half of the haloes
contain about 6 clouds because many clouds have merged, one contains 
about 15, and the rest have about 30. At $z=5$, the total number of clouds 
used to calculate the mass spectrum  
is 30; at later redshifts, the total number is greater than 100.
Masses are calculated for each cloud.  Figure 6 shows mass
spectra for all the gas clouds at nine redshifts.  
Our data appear to flatten slightly at the low
mass end, which may be a consequence of our
resolution.  Remarkably, least
squares fits to the logarithmic data show that, at all redshifts,
a power law fits the data very well for masses $\ge 10^{7.8} M_{\sun} $, 
with $1.6 \le
\alpha \le 1.9$.  There is, at best, only very weak evidence for an evolutionary
trend for $\alpha$ decreasing over time.  The figure indicates that the
original low mass clouds accrete mass or merge with other clouds,
moving to the right in the panels over time.  The low mass clouds
are continually replenished as new density condensations grow.  The
low mass end of the spectrum is not depleted but contains essentially
the same number of clouds at all redshifts.  Thus, the power law slope
is expected to be steeper at high redshift, before high mass clouds
have had time to build up, but flattens over time as mass moves to the
right in the figure. Because clouds are simply
agglomerating without being shredded by feedback from
internal star formation (the latter effect was included
in MP's model), the upper mass end of
the spectrum continues to be pushed to higher masses.
It is likely that a correct treatment of star formation within
these clouds would result in the truncation of this mass spectrum.
At this stage, therefore, the predictions of the analytic HP model
and its numerical extension (MP) are seen to arise very
naturally within a hierarchical cosmological model.
We discuss the ramifications of this in \S4.

\begin{figure*}
\centering
\centerline {\epsfxsize=15.0cm \epsfbox[18 144 592 718]
{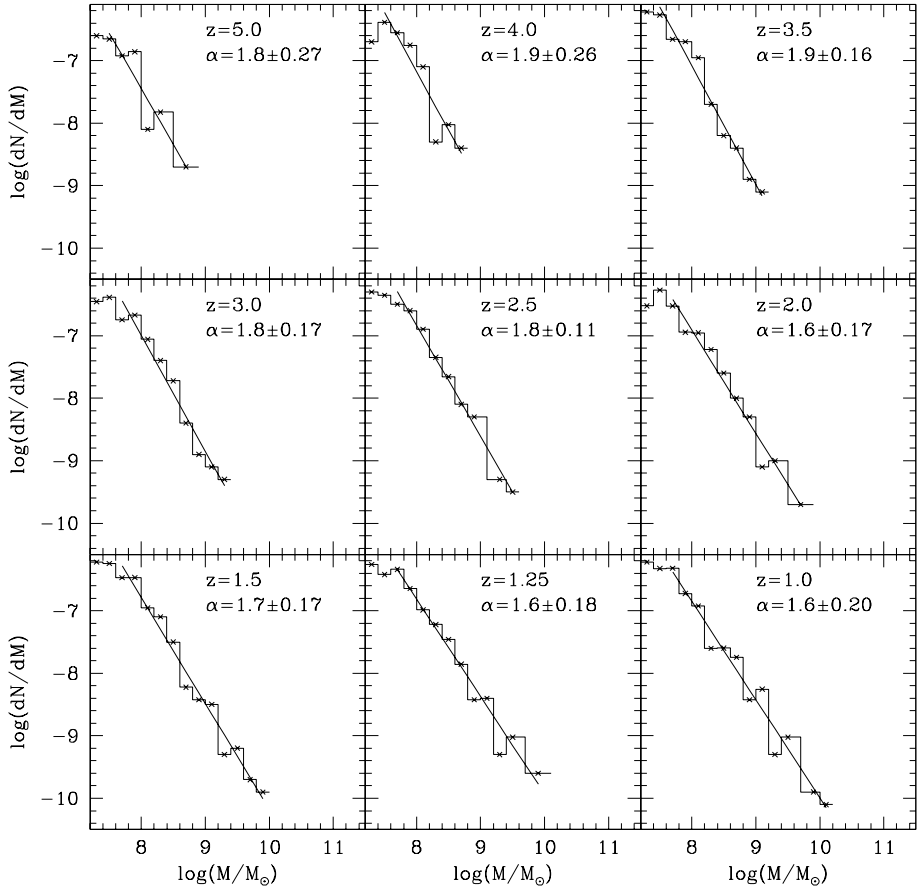}}
\caption{Mass spectrum of combined gas clouds for all
seven runs shown at nine redshifts.  The solid lines show the
power-law fits of eq. (1).  The slopes, $\alpha=\Delta {\rm log}
(dN/dM)/\Delta {\rm log} M$, are shown at the top left, below the
redshift.
}
\label{fig:figure5}
\end{figure*}

Figure 7 shows the mass spectra for the dark matter in the bound objects. 
The solid line is the least squares fit.
Again, the GCS power law fits well, with $1.5 \le \alpha \le 1.8$.
The evolutionary trend of the slope is weaker here.  If the trend
is cosmological, one explanation is suggested by Figures 2 and 3.
The dark matter forms potential wells into which gas falls, as seen by
comparing the number and density of clumps at any one early
redshift, say $z=3$, in the two figures.  Once the dark matter clumps are
formed, they do not increase in mass as the associated gas does.  Figure 7 
shows that the mass does not evolve strongly to the right over
time, but that most high mass dark matter clumps are in place by $z=3$.
The flattening or turnover at small mass is even more apparent for the
dark matter than the gas spectra.  This turnover also occurs in the globular
cluster systems of real galaxies (MP).  In observed GCS, it is difficult to
determine whether the change in the power law slope is physical
or due to incomplete sampling at low masses.  In Figure 7, it appears that
a collection of small dark matter clumps, with masses less than $10^9 M_{\sun}$,
form at intermediate redshifts but higher resolution simulations are needed
to determine if the turnover is physical or due to limited resolution.
Meanwhile, we limit the fit to the mass spectra to clumps with masses
greater than $3 \times 10^8 M_{\sun}$.
 
The dark matter mass spectra in Figure 7 are also
compared to the Press-Schechter linear analytic description of the
evolution of mass in a hierarchical cosmology (Press \& Schechter 1974).
The Press-Schechter multiplicity function predicts how
bound objects condense out of a density field described by the
cosmological power spectrum $\vert \delta_k^2 \vert \propto k^n$,
\begin{equation}
 N(M,z)dM \propto M^{n/6-3/2} exp\left[-\left({M \over
M_*(z)}\right)^{n+3 \over 3}\right]dM
\end{equation}
where $M_*(z) \propto (1+z)^{(-6/(n+3))}$.  The power spectrum of
our cosmological simulations is a scale invariant spectrum with $n=1$
but extrapolated to smaller scales (Efstathiou, Bond, \& White 1992).  Figure 7 
shows fits using a non-linear Marquardt method to masses $M
> 3.5 \times 10^8 M_{\sun}$. The dotted and dashed lines are for
$n=1$ and $n=0$, respectively.

\begin{figure*}
\centering
\centerline {\epsfxsize=15.0cm \epsfbox[18 144 592 718]
{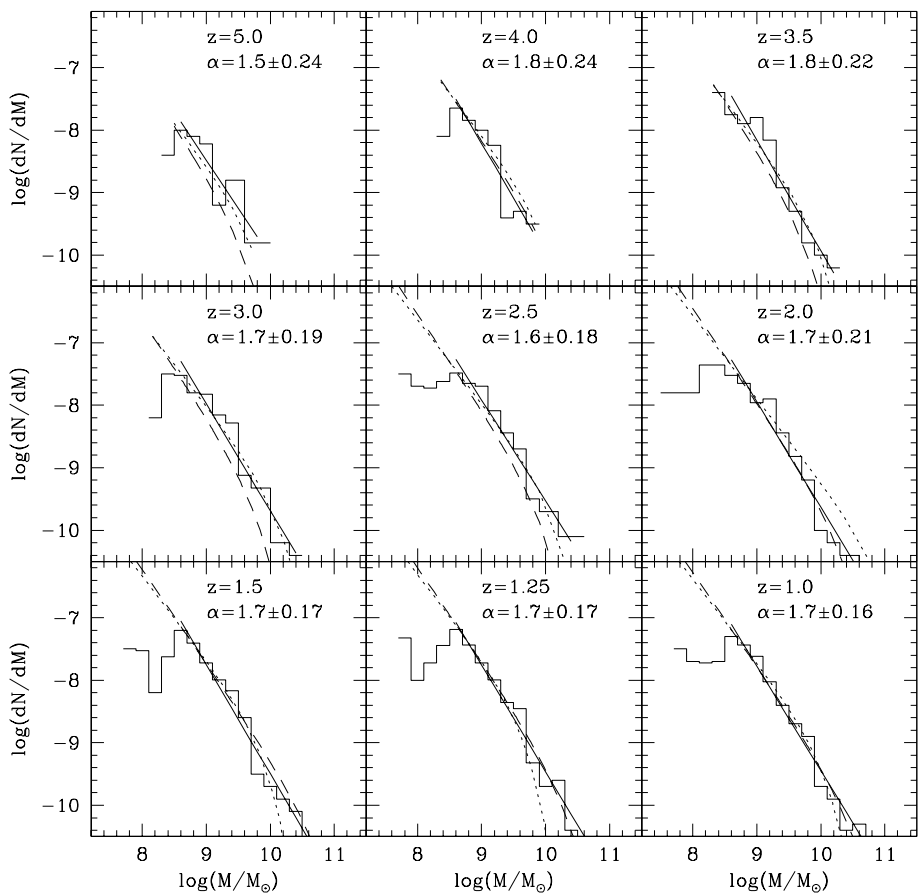}}
\caption{Mass spectrum of combined dark matter
clumps for all seven
runs shown at nine redshifts.  The solid lines show
power-law fits, as described in the previous figure caption.
The dotted and dashed lines are fits of the
Press-Schechter multiplicity function of eq. (1) with $n=1$ and $n=0$,
respectively.
}
\label{fig:figure6}
\end{figure*}

The Press-Schechter method is limited by the
assumption of linear growth of bound objects but can be extended
to treat non-linear epochs of collapse and merging (\eg Bower 1991,
Bond \etal 1991) Application of the original Press-Schechter mass
function to our data shows that the fit diverges at high masses where the
exponential part of equation (4) is important.  In addition, the
reliance on redshift in the exponential is weak in the data.  Below the
exponential cutoff, the fit is a simple power law.  However, for
$n=1$, the expected slope is $-1.3$, flatter than the least
squares fit slopes. That agglomeration-produced, power-law models fit the dark matter haloes better than Press-Schechter indicates that the effects of merging
overwhelm those of linear growth, as expected in hierarchical cosmologies.

\subsection{ Individual Cloud Properties}

Whereas the mass spectrum predicted by observations of globular
cluster systems is fit at all redshifts by the supergiant star-forming
clouds assumed to be their
birthplaces, individual properties of clouds are partially
determined by the resolution of the simulations.  The HP SGMC properties are
based on the application of the virial theorem.
They predicted that  median SGMC in the power-law spectrum
would have a mass $\sim 7 \times 10^8 M_{\sun}$,
a radius $r \sim 900 pc$, surface density $\Sigma \sim 260 M_{\sun}/pc^2$,
and density $\rho \sim 0.22 M_{\sun}/pc^3$. Figure 6 shows that
the low mass end fit by the power law is populated by clouds of the
expected mass, a few times $10^8$ to a few times $10^9$.

The assumptions of approximate virial equilibrium for the SGMCs can be
tested by noting that the virial theorem for uniform spherical
self-gravitating objects predicts that a cloud's mass
is related to its radius and internal (1-D) velocity dispersion
by $ M = 5 \sigma^2 r / G $.  The state
of our clouds is tested by plotting the virial theorem in the form
$M $  versus the quantity $  \sigma^2 r$ for the numerical data.
Figure 8 shows the logarithmic expression of the
quantities for the clouds of Runs 1 --7 at nine redshifts.  Least
squares fits give slopes, $a$, close to unity, indicating that, as a
whole, cloud ensembles are indeed near virial equilibrium.
However, Figure 8 also shows that
some individual clouds are not virialized but are just condensing
out of the primordial density perturbation field or affected by recent
interactions (objects of higher than average $\sigma^2 r$ with  
masses around $ 10^8 M_{\sun}$ as an example).

\begin{figure*}
\centering
\centerline {\epsfxsize=15.0cm \epsfbox[18 144 592 718]
{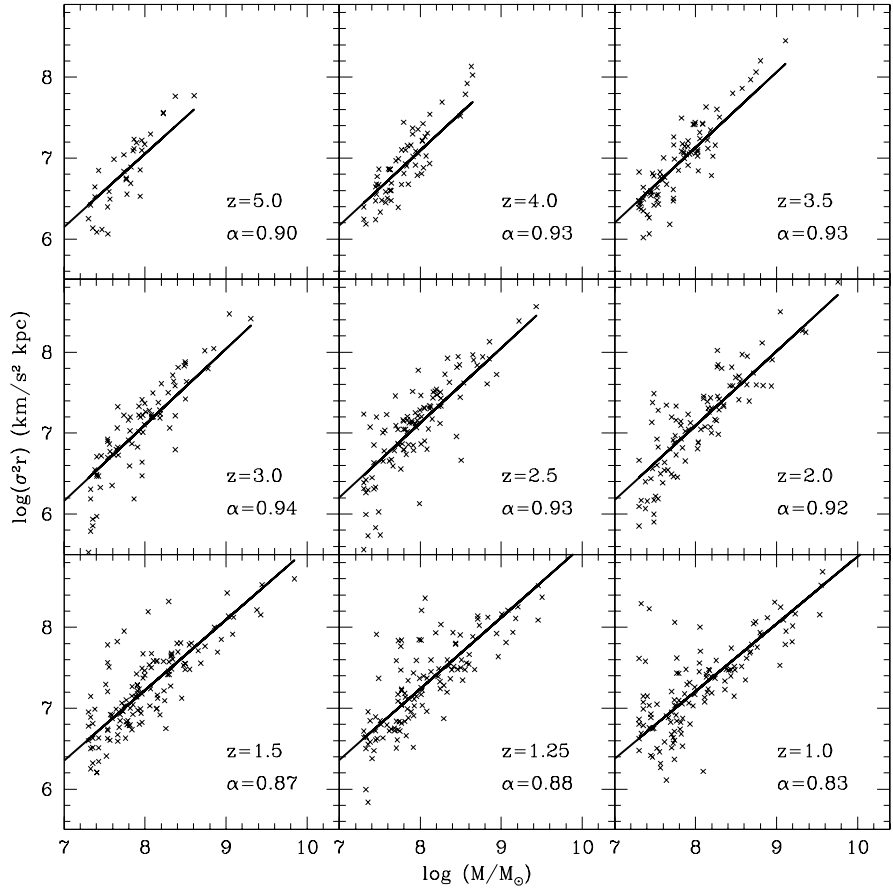}}
\caption{Virial theorem in form $M \propto
\sigma^2 r$ for Runs 1 -- 7 gas clouds.  Redshifts and least squares fits are
shown at the top right in each panel.
}
\label{fig:figure7}
\end{figure*}

With gas particle masses of a few times $10^6 M_{\sun}$, the
length resolution in the simulations is limited to $\approx 1$kpc.
Because the gravitational potential in the simulations is softened in order to
avoid two-body effects, the radii of the clouds do not evolve with
their mass. Over three decades in mass and between redshifts $z=5$ and $z=1$,
cloud radii range from $0.5 \le r \le 5$ kpc.  Due to resolution limits,
a typical cloud is approximately twice the predicted size.  Thus,
the typical surface density it a factor of four too low and the typical
density a factor of nine too low to be compared directly to SGMCs.

Any model of hierarchical structure formation that
contains some prescription for cooling the gas produces an interstellar medium
that has both cooler, self-gravitating
clouds and a hotter, diffuse medium.  The diffuse ISM
provides the pressure that truncates the sizes of molecular clouds
in our own and other galaxies. Similarly, 
SGMCs are assumed to be isothermal clouds
confined by a surface pressure due to a hotter external medium (\eg HP).

The temperature structure of the
star-forming clouds in Runs 1 -- 7 cannot explicitly represent
that of the interstellar medium because star formation, feedback, and other
methods of heating the ISM are not treated in our simulations.  Nevertheless,
the gas has a temperature structure at $z=1$ in which the temperature of
particles in the centers of the clouds is 
$\le 10,000$K, the immediately surrounding material is between $10,000$K
and $12,000$K, around which particles of $>12,000$K are smoothly
distributed. As expected,
the centers of the clouds, where star formation should occur in 
cool, dense cores,
contain low temperature gas, while the more diffuse surrounding medium is
hotter.
Although not fully representative of the multiphase ISM, the temperature
structure is that of a clumpy, warm, neutral medium surrounded by a more
diffuse, warm, ionized medium. 

\section{Discussion and Summary}

The simulations clearly establish that, in $\tau$CDM cosmological
models at least, the hierarchical growth of dark matter haloes
is accompanied by the growth of self-gravitating, supergiant
clouds through the related processes of cloud-cloud collisions
and agglomeration.  The mass spectrum that we find for SGMCs
takes the expected power-law form.  That clouds are
more concentrated than the dark haloes indicates that 
ISM properties are the key to cloud production.  Our SGMCs are not merely
condensations in the centers of mini-haloes.  Rather, they
are self-gravitating entities that are similar to very massive GMCs that
form in starburst galaxies today.

We have good indications that these results pertain to other
cosmological models as well.
We evolved one halo from the $\Lambda$CDM model mentioned in
\S 2, and found similar
mass spectra and other features of the SGMCs reported here.
The generality of our results is, in fact, to
be expected on physical grounds
because, regardless of the hierarchical model for the dark matter,
the gas dynamical processes remain the same.
Gas will collapse into the haloes, agglomerate there,
and agglomerate with the gas in other nearby haloes that
merge with their hosts.  Gas agglomeration is a robust
process that has been studied
for decades in many ISM contexts (\eg reviews by Elmegreen 1993,
Pudritz 2000), 
beginning with Oort's famous
model for gas clouds in the ISM. In a future paper, results of a 
study of gas physics and SGMC formation in $\Lambda$CDM 
models will be reported.

It has often been suggested that star
formation within assembling galaxies is restricted to the disks of
spiral-like galaxies.  
Thus, it is interesting to observe that the spatial distributions of SGMCs are
{\it not} disk-like in any of Runs 1-- 7, but rather are spread out in
an irregular fashion throughout the still-assembling dark matter
haloes that, at $z=0$, will be the haloes of galaxies with masses of a
few $\times 10^{11}$ M$_{\sun}$.  If SGMCs give rise to star and
globular clusters, the clusters are likely to be distributed throughout the
halo of the galaxy at our epoch.  

Several models assume that
globular clusters form by fragmenting out of massive, self-gravitating
sheets of gas.  These sheets could result from cloud-cloud collisions
(\eg Kumai \etal 1993) or swept-up supershells
from earlier epochs of primordial star formation (\eg Brown \etal 1995).
Collisions between clouds, it has been suggested, could compress the gas
into thin, dense, post-shock sheets that may only be 10s of parsecs thick.
Gravitational fragmentation is then invoked to form clusters by gravitational
instability in such self-gravitating layers.
Our simulations
do not suggest that our SGMCs are significantly crushed  by
collisions although we hasten to
emphasize that our gas smoothing parameter would not allow us
to resolve any layer thinner than 0.6 kpc.
The central point remains:
we find that SGMCs are assembled in lower mass potential wells
systematically with time, as Figure 6 shows.
Cloud-cloud collisions in lower mass dark haloes
favors their agglomeration, rather than
their disruption, because the cloud-cloud velocity dispersion
in lower mass haloes is significantly lower than it is in more
massive galaxies that emerge later.
Secondly, we note that a major prediction of sheet fragmentation
is that there is a preferred mass
scale that corresponds to the most rapidly growing, gravitationally
unstable mode in the sheet (\eg Elmegreen \& Elmegreen 1979).
The expected, strongly peaked mass spectrum that results from
this process is at odds with the
observed, power-law mass spectrum of clumps and
globular clusters.
We suggest that sufficient mass, which can
be assembled by a variety of different physical processes -- and
not external triggers -- is all that
may be required to form clusters in the ISM of the Milky Way, as well as
in assembling galaxies at high redshift (see \eg HP).

Observations of star formation
in current GMCs indicate that the typical star forms as a member of a
star cluster.  Most star clusters formed in the molecular clouds of
our galaxy today become unbound as the natal clumps in their host GMCs
are dispersed.  The same process may occur for cluster formation in
SGMCs.  While we anticipate that SGMC internal clumps will be the
sites for all of the star formation in the assembling galactic halo,
only a few of the putative clusters will be sufficiently bound to
survive clump dispersal.  Bound clusters become globular
clusters. Those that do not remain bound -- because of lower star
formation efficiency within their natal clumps -- will disperse (as
the open clusters in our own Milky Way do today) and produce a halo
star population.  It is likely that these clouds also produce the
stars that account for the very faint ($28 \hbox{mag arcsec}^{-2}$),
stellar luminosity observed at large radii around many nearby galaxies
(Malin \& Hadley 1999).  Proof of these assertions must await further
simulations that can include reasonable rules for star formation.

Which aspects of the formation and evolution of star-forming clouds
are affected by the underlying cosmological model?
The most obvious is that the cosmological model dictates the
rate at which the dark matter haloes evolve. This, in
turn, determines when the gas undergoes substantial cooling and agglomeration.
Thus, the model determines the time at which objects such
as globular clusters will be produced. The measured ages of the globular clusters
can provide an important constraint on any
simulation of star formation during galaxy formation.
By requiring that cosmological models
form SGMCs and, presumably, globular clusters within the
age constraints provided by observed globular clusters, 
the field of viable theories is further narrowed.

Consider the case of our
$\tau$CDM simulations.  The earliest time
at which SGMCs in the right mass range begin to
appear in the simulations is a redshift of order $z \simeq 8$.
This is an approximate result; yet higher resolution
is needed to determine what occurs at higher
redshifts.  The oldest that a globular cluster could be in this
cosmology is $\sim 10$ Gyr if globular cluster
formation began at $z= 8$.
This is at odds with the latest calculations of globular cluster
ages.  If the discrepancies that result from distance scale
uncertainties are considered, the upper limit of Milky Way
globular cluster ages can range anywhere from 12 to
17 Gyr (Grundahl \etal 2000).  By invoking
further constraints, these authors suggest an age of 16 Gyr
for the oldest cluster.  Even with this large latitude,
$\tau$CDM models with representative Hubble constants 
cannot account for globular cluster ages.
On the other hand, $\Lambda$CDM models, with their longer
ages due to accelerating expansion, likely provide a reasonably
comfortable age to admit globular clusters.

Nevertheless, the lack of feedback from star formation on the process of
SGMC assembly is an important limitation.
A steep drop in the mass spectrum
of globular clusters above several million solar masses
is interpreted by HP as arising from the destruction
of SGMCs through supernova activity.  A steady-state
solution to the agglomeration/fragmentation equations
arises if cloud growth is balanced by the destruction
of the most massive clouds through disruption by
star formation (eg. Oort 1954,
Kwan 1979).
When the time-scale for forming a yet more massive cloud
becomes longer than the cloud's destruction time due
to massive stars and supernovae within it,
a strong break in the mass spectrum along with a sharp
reduction in the number of clouds is seen. 
  
A steady state mass spectrum for the SGMCs 
does not occur in our models because
star formation is ignored.  Thus, our SGMCs establish
the correct power-law form, and continue to grow
to ever larger masses throughout the simulation.  The
disruption of SGMCs must 
occur at the very latest by $z = 1$, when all indications are that
a great deal of star formation must occur.
However, globular cluster data suggest that this truncation occurred
at higher redshifts when supernovae ripped
their natal SGMCs apart (MP).
Prescription of a correct rule for star formation will inevitably
truncate the mass spectrum of SGMCs.

The fundamental limitation of the simulations presented in 
this paper is that we are unable to resolve the substructure
of the SGMCs.  Star formation will take place in the clumps
within our clouds, but the expected masses of the clumps
is less than the minimum mass of our simulation particle.
As we have made abundantly clear however, it is hazardous
to blindly apply star formation rules to the bulk of the gas
in an SGMC.  An analysis of the onset of star formation
must await a future, much higher resolution simulation.  
Nevertheless, we have clearly shown that we can resolve
and study the SGMCs which provide the arenas in which
clump, and cluster formation, are likely to take place.

A limitation to all current simulations is that the
support of clouds by magnetic pressure, as well as by
MHD turbulence of some kind, is not modeled in the purely
hydrodynamic codes.  This is not as irrelevant as
might first appear.  Any local molecular cloud undergoes rapid cooling as
it forms dues to the efficiency of CO cooling.
But, as already noted, this does not lead to catastrophic rates of star formation
(\ie within a dynamical time) throughout the cloud.  
Catastrophic star formation is prevented because of the effective support provided
by MHD turbulence and magnetic pressure within the clouds. Thermal pressure
becomes irrelevant. Observations indicate that dynamically significant
magnetic fields are present in high redshift galaxies (Kronberg 1994) and,
therefore, are an important factor in the
physics of any self-gravitating clouds found within them.
Magnetic fields could be generated in the turbulent,
colliding gas in the assembling dark haloes by localized dynamo
processes (\eg Pudritz \& Silk 1989).
Magnetic fields of the required strength (several
$\mu$ Gauss) can predate the formation of the
first stars and should also be addressed within
improved numerical models.

In conclusion, we have found that the supergiant self-gravitating clouds
(SGMCs) do indeed form in the dark matter haloes in our $\tau$CDM 
simulations.  The SGMCs
assemble through cloud-cloud collisions as gas collects both by infall
into haloes and through the constant merger of gas-containing haloes.
A robust power-law mass spectrum for the SGMCs is found in our
simulations, 
wherein, in the mass range from $\approx 10^7 - 10^{10} M_{\sun}$, 
$dN/dM \propto M^{1.7 \pm 0.2}$.
This is exactly the spectrum that is predicted to
account for the formation of globular cluster systems around all types
of galaxies (HP).  Further independent support for our conclusions comes
from studies of chemical enrichment in hierarchical formation of the
galactic spheroid, in which an excellent match to the observed 
metallicity distributions of globular clusters in possible provided that
their progenitor clouds have a mass spectrum of the same form as our
own SGMC spectrum (Cot\'e \etal 2000).

The properties of the individual SGMCs were also determined in our study.
The clouds appear to be in near virial
equilibrium. Higher resolution simulations will ultimately provide
better determinations of their properties, as well as 
resolve the substructure of clumps that are expected to be the
birthsites of stars and star clusters in the early universe.  
Finally, preliminary
indications are that SGMCs with the same mass spectrum, and formed
by similar processes, occur in $\Lambda$CDM models.  These results
will be studied and reported in a future paper.

\section*{Acknowledgments}
MLW and REP are especially grateful to Vincent Eke for generously
providing initial conditions from his dark matter simulations, without
which this project would not have been possible.  
We thank Bill Harris for interesting discussions and for comments
on a draft of our manuscript.
REP is indebted to Simon
White and the Max-Planck-Institut f\"ur Astrophysik for the interest
and generous support that he enjoyed during a Research Leave in 1997,
when this work was begun.  
MLW acknowledges
the support of a McMaster postdoctoral fellowship from the Dean
of Science, Peter Sutherland, and a PSC-CUNY Research Award.  
The research of REP is supported by grants from the Natural
Sciences and Engineering Research Council (NSERC) of Canada.

\end {document}